\documentclass[superscriptaddress,twocolumn,showpacs,
preprintnumbers,amsmath,amssymb,nofootinbib,
longbibliography,aps,prd,10pt]{revtex4-1}
\usepackage{graphicx}

\usepackage{color}

\usepackage{amsmath}
\usepackage{amssymb}
\usepackage{amsfonts}
\usepackage[usenames,dvipsnames]{xcolor}
\usepackage{graphicx}  
\usepackage{enumerate}
\usepackage{soul}
\usepackage{wasysym}

\usepackage[colorlinks=true,
			citecolor=green,
			linkcolor=red,
			filecolor=cyan,
			urlcolor=magenta,
			backref=false]{hyperref}

\newcommand{\hh}{,\hspace{0.5cm}}
\newcommand{\hhh}{,\hspace{0.2cm}}

\newcommand{\val}[1]{#1}
\newcommand{\jen}[1]{#1}

\newcommand{\ts}[1]{{\boldsymbol{#1}}}         
\newcommand{\lap}{\triangle}
\newcommand{\hor}{\stackrel{H}{=}}

\newcommand{\be}{\begin{equation}}             
\newcommand{\ee}{\end{equation}}               
\newcommand{\ba}{\begin{eqnarray}}             
\newcommand{\ea}{\end{eqnarray}}               
\newcommand{\n}[1]{\label{#1}}

\begin{document}

\title{Stationary black holes with stringy \jen{hair}}

\author{Jens Boos}

\email{boos@ualberta.ca}

\author{Valeri P. Frolov}

\email{vfrolov@ualberta.ca}

\affiliation{Theoretical Physics Institute, University of Alberta, Edmonton, Alberta, Canada T6G 2E1}

\date{\today}

\begin{abstract}
We discuss properties of black holes which are pierced by special configurations of cosmic strings. For static black holes we consider radial strings in the limit when the number of strings grows to infinity while the tension of each single string tends to zero. In a properly taken limit the stress-energy tensor of the string distribution is finite. We call such matter stringy matter. We present a solution of the Einstein equations for an electrically charged static black hole with the stringy matter, with and without cosmological constant. This solution is a warped product of two metrics. One of them is a deformed two-sphere whose Gaussian curvature is determined by the energy-density of the stringy matter. We discuss the embedding of a corresponding distorted sphere into a three-dimensional Euclidean space and formulate consistency conditions. We also found a relation between the square of the Weyl tensor invariant of the four dimensional spacetime of the stringy black holes and the energy density of the stringy matter. In the second part of the paper, we discuss test stationary strings in the Kerr geometry and in its Kerr--NUT--(A)dS generalizations. Explicit solutions for strings that are regular at the event horizon are obtained. Using these solutions the stress-energy tensor of the stringy matter in these geometries is calculated. Extraction of the angular momentum from rotating black holes by such strings is also discussed.
\end{abstract}

\pacs{04.20.Jb, 04.70.Bw \hfill Alberta-Thy-13-17}

\maketitle

\section{Introduction}
Cosmic strings are one-dimensional topological defects which might be formed in the symmetry breaking phase transitions in the Early Universe \cite{vilenkin2000cosmic}. The string's tension $\mu_s$ and its width $\eta_s$ are \jen{related to the} characteristic energy scale of the corresponding phase transition $m$ \jen{via}
\be
\mu_s\sim {m_{Pl}\over l_{Pl}}\left( {m\over m_{Pl}}\right)^3\hh \eta_s\sim l_{Pl} {m\over m_{Pl}}\, .
\ee
Cosmic strings formed in the Early Universe would distort the cosmological microwave background (CMB). However, the observations indicate that their contribution to CMB cannot be more than $10\%$ \cite{Ade:2013xla}.

The dynamics of a test string in an external gravitational field $g_{\mu\nu}$ is described by the Nambu--Goto action
\be
I=-\mu_s \int d^2\zeta \sqrt{- \det\left(\gamma_{ab}\right)}\, ,\ \
\gamma_{ab}= g_{\mu\nu}\partial_{\zeta^a} x^{\mu}\partial_{\zeta^b} x^{\nu}\, .
\ee
$\mu_s$ is the string tension and $\zeta^a$ $ (a=0,1)$ are coordinates on the string world-sheet. The functions $x^{\mu}(\zeta^a)$ determine string's embedding in the bulk spacetime. The stress-energy tensor of the string is localized on its surface and is of the form \cite{vilenkin2000cosmic}
\be\n{Tmn}
T^{\mu\nu}={\mu_s\over \sqrt{-g}} \int d^2\zeta \sqrt{-\gamma} \gamma^{ab} x^{\mu}_{\ ,a}
x^{\nu}_{\ ,b} \delta^{(4)}(x^{\sigma}-x^{\sigma}(\zeta^a))\, .
\ee
For a straight string in $z$-direction in a flat spacetime with Cartesian coordinates $(t,z,x,y)$ it takes the form
\be
T_{\mu}{}^{\nu}=\mbox{diag}(-\mu,-\mu,0,0)\, , \ \  \mu=\mu_s \delta(x)\delta(y)\, .
\ee
The spacetime is locally flat outside the string and it has the angle deficit $\nu=8\pi \mu_s$. Such a space can be obtained by  cutting out a wedge of angle $\nu$ along along $z-$axis and gluing together the edges \cite{vilenkin2000cosmic,Puntigam:1996vy}.

Several interesting effects occur in a situation when a cosmic string, passing near a black hole, is caught by the latter (see e.g.\ the book \cite{anderson2015mathematical} and references therein).
The simplest case corresponds to an infinitely long straight string piercing a black hole \cite{Aryal:1986sz}. Properties of static black holes pierced by a polyhedral set of radial straight strings were discussed in \cite{Dowker:1992xq,Frolov:2001xra}. In the paper \cite{Frolov:2001uf} the authors introduce a notion of a ``thorny sphere,'' which is  everywhere locally isometric to a round two dimensional sphere except at a finite number of isolated points where it has conical singularities. Using thorny spheres a general solution for a black hole pierced by an arbitrary number of radial strings was constructed. Such configurations can be used for quantum mining of energy from black holes \cite{Lawrence:1993sg,Frolov:2000kx}. Such a model was discussed in connection with the information loss paradox \cite{Giddings:2014ova,Giddings:2017mym}.
The interaction of classical strings with a rotating black hole can be also be used for extraction of the energy from the latter \cite{PhysRevD.54.5093,0264-9381-14-5-015,1402-4896-62-2-3-005,Kinoshita:2016lqd}. In some aspects this process is similar to the Blandford--Znajek mechanism \cite{1402-4896-65-1-002,Kinoshita:2017mio}.

In the present paper we continue the study of interaction of cosmic strings with black holes. We generalize the results on the thorny black holes to the case when the number of strings attached to the black hole grows to infinity while their tension decreases. In the properly chosen limit \jen{(``smearing the string'')} such a configuration describes a radial distribution of what is called the stringy matter. We discuss properties of static and stationary black hole with such stringy-matter hair.

The paper is organized as follows. After the Introduction  we review geometrical properties of the spacetime for a straight smeared string (Section~II). In Section~III we describe a solution of the Einstein equations for a static charged (A)dS black hole with the stringy matter. The corresponding metric is a warped product of a 2D metric of the $(t,r)$ sector of the unperturbed solution and a 2D metric of the distorted sphere, the Gaussian curvature of which is determined by the stress-energy tensor of the stringy matter. The embedding of a distorted sphere into 3D Euclidean space, consistency conditions, and the relation between its 2D Gaussian curvature and 4D curvature invariants are also discussed in this section. Single strings piercing a Kerr black hole and the stress-energy tensor of corresponding stringy matter are discussed in Section~IV. Section~V contains the generalization of these results to the case of a Kerr--NUT--(A)dS black hole. Section~VI contains a brief discussion of the obtained results.

\section{Geometry of a straight smeared string}

Since the force between any two straight cosmic strings with arbitrary orientations vanishes, one can consider a static ensemble of such cosmic strings which are in passive equilibrium. In particular, one can choose a set of parallel cosmic strings and take a limit of their continuous distribution. In such a limit the number of strings grows to infinity, while the tension of each individual string decreases. An action for such stringy matter was discussed in \cite{Ivanov:2001wb,Ivanov:2003sq}. The gravitational field for such one-dimensional straight distribution of the stringy matter can be written as \cite{vilenkin2000cosmic}
\be \n{i.1}
ds^2=-dt^2+dz^2+d\rho^2+f^2(\rho,\phi) d\phi^2\, .
\ee
Substituting the metric \eqref{i.1} in $(t,z,\rho,\phi)$ coordinates into the Einstein equations
\be
G_{\mu}{}^\nu\equiv R_{\mu}{}^\nu-{1\over 2} \delta_{\mu}^{\nu} R=8\pi T_{\mu}{}^\nu\, ,
\ee
one finds
\be
T_{\mu}{}^{\nu}=\mbox{diag}(\mu,\mu,0,0)\, ,\ \  \mu=\mu(\rho,\phi)=-{1\over 8\pi} f^{-1} \partial_{\rho}^2 f\, .
\ee
This is exactly what one would expect for a one-dimensional stringy matter distribution.

We assume that the angle coordinate $\phi$ has the period $2\pi$. Then the regularity of the metric (\ref{i.1}) at  $\rho=0$ implies that $f\sim \rho$ near this point. 

In the domain outside the matter the Gaussian curvature vanishes and the geometry can be embedded into three-dimensional Euclidan space as a cylinder \cite{massey1962}. The metric in this domain can be written in the form (\ref{i.1}) with $f=b\rho$. The parameter $b$ is connected to the angle deficit via $\nu=2\pi(1-b)$.

The tension $\mu$ of the stringy matter is directly related to a special geometric invariant. Namely, consider the two-dimensional metric 
\be
\jen{ds_{(2)}^2}=d\rho^2+f^2(\rho,\phi) d\phi^2\, .
\ee
The Gaussian curvature $K$ of this 2D metric, which is connected to the 2D Ricci scalar ${}^{(2)}R$ via $K={}^{(2)}R/2$, is
\be\n{a.3}
K={1\over 2}{}^{(2)}R = - f^{-1} \partial_{\rho}^2 f\, .
\ee
Thus one has
\be
K=8\pi \mu\, .
\ee

\section{Static black holes with stringy hair}

\subsection{Geometry}

Our starting point for constructing a solution for a static black hole with stringy hair is the following metric:
\ba\n{met}
&& ds_0^2=-f dt^2+{dr^2\over f}+r^2 d\omega_0^2\, ,\\
&& d\omega_0^2=d\theta^2+\sin^2\theta d\phi^2\, .
\ea
For 
\be\n{fff}
f=1-{2M\over r}+{\jen{Q}^2\over r^2}-{1\over 3} \Lambda r^2 \, ,
\ee
this metric describes a static charged spherically symmetric black hole in an asymptotically (A)dS spacetime, see also \cite{Moskalets:2014hoa}. It is a solution of the Einstein--Maxwell equations
\ba
&&G_{\mu}{}^\nu\equiv R_{\mu}{}^\nu-{1\over 2}  \delta_{\mu}^{\nu} R=8\pi T_{\mu}{}^\nu\, ,\n{EE}\\
&&F^{\mu \nu}{}_{;\nu}=0\hh F_{\mu\nu}=2A_{[\nu ,\mu]}\, ,
\ea
with the potential
\be \n{AAA}
A_{\mu}=(\jen{Q}/r)\delta^t_{\mu} \, .
\ee
The stress-energy tensor which enters the Einstein equation is
\be \n{TTT}
T_{\mu\nu}=-{1\over 8\pi}\Lambda g_{\mu\nu} +T^{(em)}_{\mu\nu}\, .
\ee
Here the first term in the right-hand side is just a cosmological constant while the second term is the stress-energy tensor of the static electric field
\ba
&&T^{(em)}_{\mu\nu}={1\over 4\pi}(F_{\mu\beta}F_{\nu}^{\ \ \beta}-{1\over 4} g_{\mu\nu} F_{\alpha\beta}F^{\alpha \beta})\nonumber \, , \\
&&\jen{g{}^{\nu\rho} T^{(em)}_{\mu\rho}}=\mbox{diag}(-Y,-Y,Y,Y)\, ,\ \  Y={\jen{Q}^2\over 8\pi r^4}\, .\n{YYY}
\ea

The corresponding black hole solution with the stringy hair is obtained by the following deformation of the metric (\ref{met})
\ba\n{dmet}
&& ds^2=-f dt^2+{dr^2\over f}+r^2 \val{d\omega^2}\, ,\\
&& d\omega^2=\exp(2\sigma) d\omega_0^2\, .
\ea
The metric is still a warped space where the 2D round sphere is distorted and possesses the metric $ d\omega^2$. Let us note that the function $f(r)$ remains the same. The radius $r_+$ of the black hole horizon, $f(r_+)=0$, and its surface gravity
\be
\kappa_H=\left.{1\over 2}{df\over dr}\right|_H\, ,
\ee
are the same as for the undistorted black hole. Introducing the advanced time 
\be
dv=dt+{dr\over f}
\ee
one can check that the metric in these coordinates,
\be
\label{vvv}
ds^2=-f dv^2-2 dv dr+r^2 d \omega^2\, ,
\ee
is regular at the \jen{future} event horizon\footnote{\val{Let us notice that this metric is a special case of a general class of Robinson--Trautman metrics \cite{Robinson:1960zzb,Robinson:1962zz,Bicak:1997ne,Stephani:2003}. The metric (\ref{vvv}) reproduces the Robinson--Trautman line element (see Ch.~28 of \cite{Stephani:2003}) 
\be\n{RT}
ds^2=2r^2 P^{-2} d\zeta d\bar{\zeta}-2 du dr -2 H du^2\hh P_{,r}=0\,  .
\ee
after the evident changes $2H\to f$, $u\to v$. The angular line element $ 2 P^{-2} d\zeta d\bar{\zeta}$ takes the form $ d \omega^2$ after the transformation $\zeta=\sqrt{2} e^{i\phi} \tan \theta/2$. In order to obtain the metric for a stringy black hole one can start with the ansatz (\ref{RT}) and obtain the metric functions $P$ and $H$ by solving the Einstein equations in presence of the stringy matter.}}. The surface area of the horizon is
\be
{\cal A}= \jen{r_+^2} \int d\theta d\phi \, \exp(2\sigma)\sin\theta\, .
\ee

It is easy to check that
\begin{itemize}
\item The potential (\ref{AAA}) is still a solution of the Maxwell equations in the distorted metric (\ref{dmet});
\item The stress-energy tensor of this Maxwell field has the same form \val{(\ref{YYY})};
\item The distorted metric (\ref{dmet}) obeys the Einstein equation (\ref{EE}) where the stress-energy tensor (\ref{TTT}) is modified by adding a term
    \ba\n{PPP}
&&{\cal T}_{\mu}^{\ \nu}=\mbox{diag}\left({ -\Phi\over 8\pi r^2},{-\Phi\over 8\pi r^2},0,0\right)\, ,\n{STR}\\
&& \Phi=\exp(-2\sigma) [1-\lap \sigma] -1\, .\n{PPh}
\ea
Here $\lap$ is the Laplace operator on the unit sphere
\be
\lap={1\over \sin \theta}\partial_{\theta}( \sin\theta \partial_{\theta})+{1\over \sin^2\theta}\partial^2_{\phi}\, .
\ee
\end{itemize}

We choose the notation for $\Phi$ so that for the matter with positive energy density the function $\Phi$ is also positive.
We call the matter with the equation of state (\ref{STR}) {\em stringy matter}.

The Gaussian curvature of the 2D metric $d\omega^2$ is connected with the 2D Ricci scalar ${}^{(2)}R$ via $K={}^{(2)}R/2$. Simple calculations give
\be\n{KKK}
K={1\over 2}{}^{(2)}R = \exp(-2\sigma)(1-\lap \sigma)\, .
\ee
For a unit round sphere, when $\sigma=0$, one has $K=1$. Substituting (\ref{KKK}) in (\ref{PPh}) gives
\be\n{KP}
K=1+\Phi\, .
\ee

\begin{figure}
  \centering
  \includegraphics[width=0.5\textwidth]{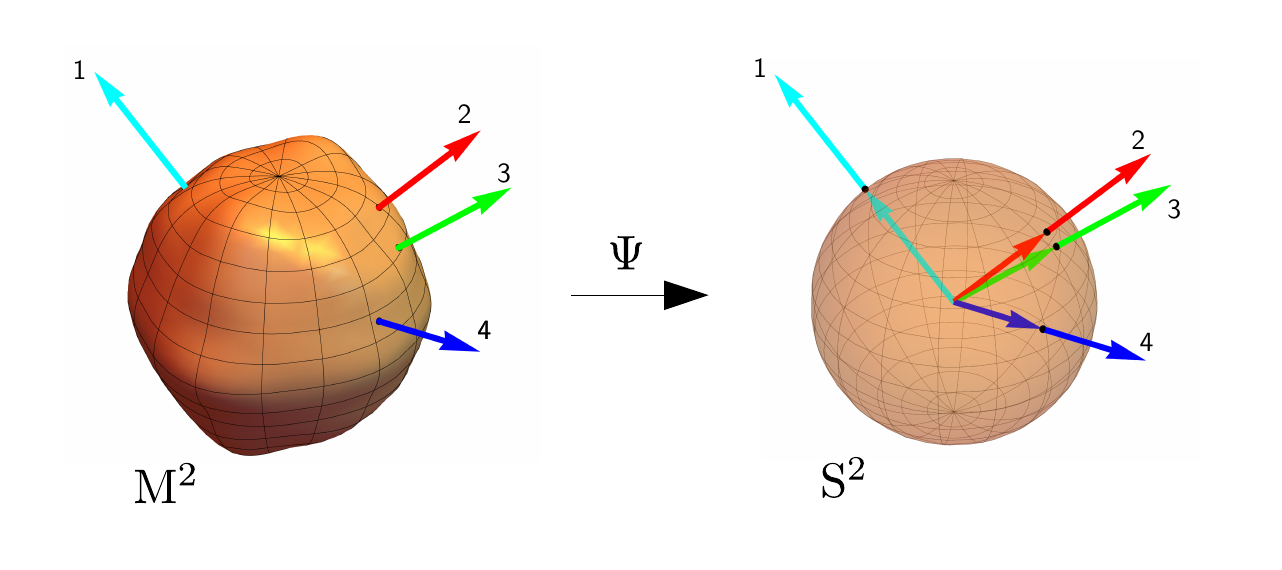}
  \caption{ Illustration of the Gauss map. This map  $\Psi: M^2 \rightarrow S^2$ transforms points on the deformed 2-sphere $M^2$ to points on the unit sphere $S^2$. }
  \label{fig:gauss-map}
\end{figure}

\subsection{Embedding. Consistency conditions.}

For positive energy density of the stringy matter $\Phi\ge 0$, so that $K\ge 1$. If one applies the results to the case of $\Phi<0$ we assume always that $\Phi>-1$. This means that $K$ is positive and, since the sphere is compact, there exist such a positive number $c$ that $K>c$. \jen{This esentially prevents the stringy matter from changing the topology of the distorted 2-sphere.} In this case the distorted sphere can be isometrically embedded into a flat three dimensional space $E^3$ as a regular surface $M^2$ \cite{pogorelov1973extrinsic}.

Let us choose a point $p$ on the deformed sphere $M^2$ and denote by $\vec{n}$ an outward unit vector normal to $M$ at this point. Let $dA$ be the element of the surface area at $p$, then we denote
\be
d\vec{A}= \vec{n} dA\, .
\ee
The Gauss map $\Psi$ of the convex surface $M^2$ to the unit round sphere $S^2$ is defined as follows. It maps a point of $M^2$ with the normal vector $\vec{n}$ to the point of $S^2$ with the same normal vector \cite{burns1991modern,banchoff1982cusps} (see Figure~\ref{fig:gauss-map}). Let us denote by $d\alpha$ and $d\vec{\alpha}=\vec{n} d\alpha$ the scalar and vector surface area elements at $\Psi(p)$. Then one has
\be
d\alpha=K dA\hh d\vec{\alpha}=K d\vec{A}\, .
\ee

The surface areas of $M^2$ and $S^2$ are
\be \n{AS}
{\cal A}=\int_{M^2} dA\hh \jen{{\cal S} = } \int_{S^2} d\alpha=4\pi\, .
\ee
The second of these relations can be rewritten as
\be
\int_{M^2} dA \ K=4\pi\, .
\ee
This is nothing but the Gauss--Bonnet theorem. Using the relations (\ref{KP}) and (\ref{AS}) one also obtains
\be
{\cal A}+\int_{M^2} dA \ \Phi=4\pi\, .
\ee
Hence, for non-negative energy density of the stringy matter, $\Phi\ge 0$, the surface area ${\cal A}$ of the distorted sphere is less than or equal to $4\pi$.

Another set of useful relations, called consistency conditions, can be obtained by using the following divergence theorem \cite{riley2006mathematical} 
\be\n{VA}
\int_V d^3 V \, \vec{\nabla}\varphi=\int_{\partial V} d \vec{A} \, \ \varphi .
\ee
Here $\partial V$ is the boundary surface restricting three dimensional volume $V$ and $d\vec{A}$ is the vector surface area element on this surface. The application of (\ref{VA}) to $M^2$ and $S^2$ for a special choice $\varphi=1$ gives
\be
\int_{M^2} d\vec{A}=0\hh \int_{S^2} d\vec{\alpha}=0\, .
\ee
The second of these relations can written in the forms 
\be
\int_{M^2} dA \, K \, \vec{n} =\int_{M^2} dA \, \Phi \, \vec{n} =0\, .
\ee
For the black hole with stringy matter the second relation has a simple interpretation: The total force acting on the black hole, which is induced by stringy matter tension, must vanish \cite{Frolov:2001uf}. It is this condition  that secures our deformed black hole solution to be static.

\subsection{Curvature invariants}

In the warped geometry (\ref{dmet}) the stringy matter directly affects the geometry of the distorted 2-sphere and its stress-energy tensor contributes to the Gaussian curvature of the latter via the relation (\ref{KP}). Let us demonstrate that the square of the Weyl tensor, characterizing the four-dimensional spacetime curvature, is also simply related to the energy density of the stringy matter. We write the metric of the stringy black hole in the form
\ba
&&ds^2 =r^2 d\hat{s}^2\hhh d\hat{s}^2=d\hat{\alpha}^2+d\omega^2\, ,\\
&& d\hat{\alpha}^2=-f r^{-2} dt^2+{dr^2\over r^2 f}\, .
\ea
We denote the quadratic invariants of the Weyl tensor for the metrics $ds^2$ and $d\hat{s}^2$ by
\be
{\cal C}^2={C}{}_{\mu\nu\rho\sigma} {C}{}^{\mu\nu\rho\sigma}\hhh
\hat{{\cal C}}^2=\hat{C}{}_{\mu\nu\rho\sigma} \hat{C}{}^{\mu\nu\rho\sigma}\, ,
\ee
respectively. Under the conformal transformation relating these metrics, \jen{$ds^2 = r^2 d\hat{s}^2$}, they transform as
\be
{\cal C}={1\over r^2} \hat{{\cal C}}\, .
\ee
Since the metric $d\hat{s}^2$ is a direct sum of two independent metrics $d\hat{\alpha}^2$ and $d\omega^2$, one has \cite{Ficken:1938}
\be
\hat{{\cal C}}={2\over \sqrt{3}}\left( {}^{(\hat{\alpha})}K + {}^{(\omega)}K \right)\, ,
\ee
where ${}^{(\hat{\alpha})}K$ and ${}^{(\omega)}K$ denote the Gaussian curvatures of the corresponding two-metrics:
\ba
&&{}^{(\hat{\alpha})}K = -\frac 12 r^2 f'' + r f' - f \, ,\\
&&{}^{(\omega)}K = e{}^{-2\sigma} \left( 1 - \Delta \sigma \right) \, .
\ea
Using (\ref{KP}) one obtains
\be
{\cal C}={2\over \sqrt{3}r^2}\left( {}^{(\hat{\alpha})}K+1 +\Phi \right)\, .
\ee
At the event horizon $r=r_+$ this relation takes the form
\be
{\cal C}_H={2\over \sqrt{3}r_+^2}\left(B +\Phi \right)\hhh
B=1+r_+ f'_+ -\frac 12 r_+^2 f_+''\, .
\ee
In the simplest case, when the charge and cosmological constant vanish, $B=3$.

\section{Rotating black hole with stringy hair}

\subsection{Principal Killing  string in the Kerr spacetime}

Let us consider a stationary string in the Kerr geometry. This metric in the Boyer--Lindquist coordinates $(t,r,\theta,\phi)$ is
\ba\n{Kerr}
&&ds^2=-\left(1-{2Mr\over \Sigma}\right) dt^2-{4Ma r \sin^2\theta\over \Sigma} dt d\phi\nonumber\\
&&+{A\sin^2\theta \over \Sigma} d\phi^2 +{\Sigma\over \Delta} dr^2 +\Sigma d\theta^2\, .
\ea
Here
\ba
&&\Delta=r^2-2Mr+a^2\hh \Sigma= r^2+a^2 \cos^2\theta\, ,\nonumber\\
&&A=(r^2+a^2)^2-\Delta a^2 \sin^2\theta\, .
\ea
This metric has two Killing vectors
\be\n{Ktp}
\ts{\xi}=\partial_t\hh \ts{\eta}=\partial_{\phi}\, .
\ee
We denote, as usual, by $r_{\pm}=M\pm\sqrt{M^2-a^2}$ the roots of $\Delta=0$. Then the event horizon, which is a null surface, is located at  $r=r_+$. The Killing vector
\be \n{nn}
\ts{n}=\ts{\xi}+\Omega \ts{\eta}\hh \Omega={a\over r_+^2+a^2} \jen{= \frac{a}{2Mr_+}}\, ,
\ee
is tangent to null geodesics which are generators of the horizon.

\begin{figure}
  \centering
  \includegraphics[width=0.5\textwidth]{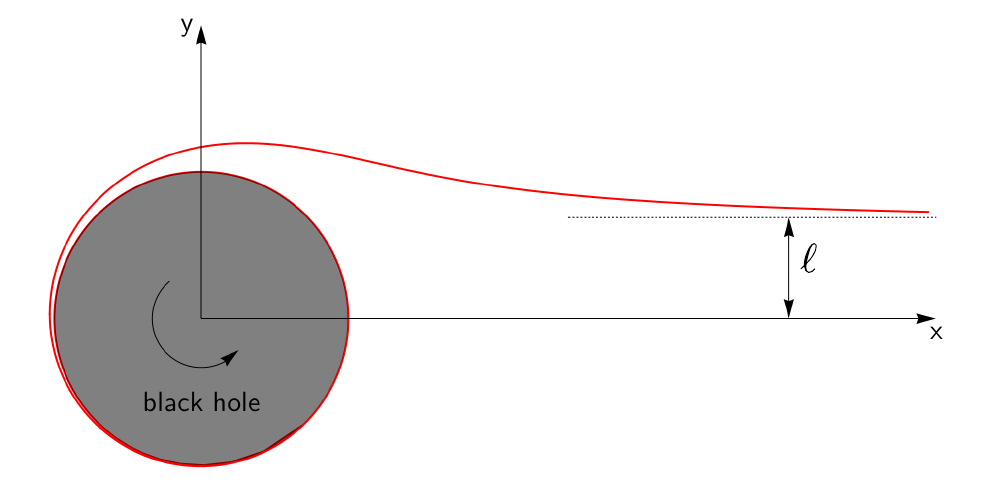}
  \caption{A line \eqref{rho}, representing the string, is shown in polar coordinates $(\rho,\phi)$ in a 2-plane.}
  \label{fig:killing-string}
\end{figure}

The stationary string equations are completely integrable in this metric \cite{Frolov:1988zn,Carter:1989bs}. There exists a special interesting solution describing a stationary string in the Kerr geometry, the world-sheet of which is a principal Killing surface. This surface has two tangent vectors. One of them is $\ts{\xi}$, while the other coincides with the null vector $\ts{l}$ tangent to a principal null geodesic.
In \cite{PhysRevD.54.5093} it was proved that the principal Killing surfaces are the only stationary timelike minimal two-surfaces that (i) cross the static limit surface, where $\ts{\xi}^2=0$, and (ii) are regular in its vicinity. \val{Such a principal Killing surface represents what we call a \emph{principal Killing string}. Such a string crosses the event horizon, and its representation in Boyer--Lindquist coordinates (for $a < M$) is
\be\n{A.1}
\phi=\phi_0+{a\over r_+ -r_-}\ln\left({r-r_-\over r-r_+}\right)\hh \theta=\theta_0=\mbox{const}\, .
\ee
The string makes an infinite number of rotations before it reaches the horizon. However, as we shall see in the next section, this is a coordinate effect connected with the choice of the angle variable $\phi$ in the Boyer--Lindquist coordinates. Let us emphasize that the dependence of the angle $\phi$ on the radius $r$ is the same for any value $\theta_0$ of the cone solution. Let us introduce dimensionless parameters
\be 
\rho=r/M\hh \alpha=a/M \hh \rho_{\pm}=1\pm\sqrt{1-\alpha^2}\, .
\ee
Then Eq.~\eqref{A.1} takes the form
\be\n{rho} 
\phi=\phi_0 +{\alpha\over 2 \sqrt{1-\alpha^2}}\ln\left({\rho-\rho_-\over \rho-\rho_+}\right)\, .
\ee
Fig.~\ref{fig:killing-string} schematically shows a line (\ref{rho}) in polar coordinates in a 2-plane.}

\subsection{Near horizon behavior}

In order to discuss the properties of the principal Killing string near the horizon we perform the following coordinate transformation:
\be
dv=dt+(r^2+a^2) {dr\over \Delta}\hh d\hat{\phi}=d\phi +a{dr\over \Delta}\, .
\ee
In these ingoing Kerr coordinates $(v,r,\theta,\hat{\phi})$, which are regular at the future horizon, the Kerr metric takes the form
\ba
&&ds^2=-{\Delta\over \Sigma} (dv-a \sin^2\theta d\hat{\phi})^2+ {\sin^2\theta\over \Sigma}\left[ (r^2+a^2)d\hat{\phi}-a dv\right]^2\nonumber\\
&& \quad \quad +\Sigma d\theta^2 +2 dr (dv-a\sin^2\theta d\hat{\phi})\, .
\ea
In these new coordinates
\be
\sqrt{-g}=\sin\theta \Sigma\, ,
\ee
and the Killing vectors (\ref{Ktp}) have the form
\be\n{Kvp}
\ts{\xi}=\partial_v\hh \ts{\eta}=\partial_{\hat{\phi}}\, .
\ee
The null generator of the horizon (\ref{nn}) is
\be
n_{\mu}\hor \left( 0,{\Sigma_+\over r_+^2+a^2},0,0\right)\, ,
\ee
where $\Sigma_+=\Sigma|_{r=r_+}$, and ``$\hor$'' denotes an equality which is valid on the horizon.

Written in the incoming null coordinates $(v,r,\theta,\hat{\phi})$ the principal Killing string equation (\ref{A.1}) takes the form
\be
\hat{\phi}=\hat{\phi_0}=\mbox{const}\hh \theta=\theta_0=\mbox{const}\, .
\ee
We use $(v,r)$ as the coordinates on the string surface. Then the induced metric is
\be
d\gamma^2={\Xi\over \Sigma} dv^2 + 2 dv dr\, ,\ \  \frac{\Xi}{\Sigma}\equiv \ts{\xi}^2=-(\Delta-a^2\sin^2\theta)\, .
\ee
Inside the ergosphere $\Xi$ is positive, so that the Killing vector $\ts{\xi}$, tangent to the string surface, is spacelike. One also has
\be
\jen{\partial{}_\gamma^2 =} \gamma^{ab}\partial_a\partial_b= 2\partial_v\partial_r-{\Xi\over \Sigma}\partial_r^2\hh
\sqrt{-\gamma}=1\, .
\ee

Using (\ref{Tmn}) one obtains
\ba\n{Tsmn}
&&T_s^{\mu\nu}=q \tau^{\mu\nu}\, ,\  \tau^{\mu\nu}={1 \over \Sigma} \left( \jen{-2\delta_v^{(\mu}\delta_r^{\nu)}}+{\Xi\over \Sigma}\delta_r^{\mu}\delta_r^{\nu}\right)\, ,\\
&&q=q(\theta,\hat{\phi}|\theta_0,\hat{\phi}_0)=\mu_s {\delta(\theta-\theta_0)\delta(\hat{\phi}-\hat{\phi}_0)\over \sin\theta}\, .
\ea
We include the subscript ``s'' in order to indicate that this is the expression valid for a single string. The horizon surface element is
\ba
&&d\sigma_{\mu}\hor -\sin\theta \Sigma_+ \delta_{\mu}^r dv d\theta d\hat{\phi}\nonumber\\
&&\quad\quad\hor -(r_+^2+a^2) n_{\mu} \sin\theta dv d\theta d\hat{\phi} \, .
\ea
Thus
\ba
&&T_s^{\mu\nu}d\sigma_{\nu}\hor  j^{\mu}  \delta(\theta-\theta_0)\delta(\hat{\phi}-\hat{\phi}_0) dv d\theta d\hat{\phi}\, ,\\
&&j^{\mu}\hor \mu_s \left( \jen{ -\delta_v^{\mu}} +{a^2 \sin^2\theta_0\over \Sigma_H} \delta_r^{\mu}\right) \,  .
\ea
It is easy to show that 
\be\n{jj}
j^{\mu}\xi_{\mu}\hor 0\, ,\ \  j^{\mu}\eta_{\mu}\hor -\mu_s a \sin^2\theta_0\,  .
\ee
To obtain the fluxes of some observable through the horizon, one has to \jen{project $T^{\mu\nu}d\sigma_{\nu}$ on} the corresponding generator and integrate the obtained scalar over $v$, $\theta$ and $\hat{\phi}$. Since the integrand does not depend on $v$, the integral over this variable give the constant $\Delta v$, which is the duration of time for which the flux is calculated. To obtain the flux per a unit of time $v$ one hence needs to divide the flux by $\Delta v$, resulting in the following flux rates of energy, $\dot{E}$, and the angular momentum, $\dot{J}$:
\be\n{EJ}
\dot{E}\hor j^{\mu}  \xi_{\mu} = 0\hhh
\dot{J}\hor j^{\mu} \eta_{\mu} = -\mu_s a \sin^2\theta_0\, .
\ee

\val{$\dot{J}$ takes its maximal (negative) value when $\theta_0=\pi/2$, that is, when the string lies in the equatorial plane. Since $a=J/M$, one obtains the following equation for the dynamics of the angular momentum of the black hole
\be 
\dot{J}=-{\mu_s\over M}J\, .
\ee
Its solution is
\be 
J=J_0 \exp(-t/t_{\mu})\, ,
\ee
where $t_{\mu}=M/\mu_s$ is the characteristic time of the deceleration of black hole rotation. We can rewrite this as
\be
t{}_\mu(M) \sim \left( \frac{M}{m_{Pl}} \right)\left( \frac{m_{Pl}}{m} \right)^3 t{}_{Pl} .
\ee
Clearly, for the electroweak scale this time scale is much larger than the age of the Universe (for both stellar mass black holes and supermassive black holes). For the Planck scale one has $t{}_\mu(M{}_{{\tiny\astrosun}}) \sim \mu\text{s}$ and $t{}_\mu(10^6M{}_{{\tiny\astrosun}}) \sim \text{s}$.
}

\subsection{Asymptotics at spatial infinity}

Let us now discuss the string properties at far distances from the black hole. In this asymptotically flat domain the Boyer--Lindquist coordinates reduce to the standard spherical coordinates. Consider sphere with radius $r$ and a point $A_0$ with spherical coordinates $(\theta_0,\phi_0)$ on this sphere. These coordinates characterize the chosen string, which at large $r$ has the following asymptotic form:
\be
\theta=\theta_0\hh \phi=\phi_0+{a\over r}+\ldots \, .
\ee
\jen{To leading order, the stress-energy tensor is
\be
T{}^{\mu\nu}_s = - \frac{q}{r^2} \left( \delta{}^\mu_t \delta{}^\nu_t + \delta{}^\mu_r \delta{}^\nu_r - \frac{2a}{r^2} \delta{}^{(\mu}_r \delta{}^{\nu)}_\phi + \dots \right) \, .
\ee
The radially inward pointing surface element is $d\sigma{}_\mu = -\delta{}^r_\mu r^2 \sin\theta \, d\theta \, d\phi$ such that one obtains for the flux vector in the limit $r \rightarrow \infty$ the following expression:
\be
j{}^\mu = \int\limits_{S^2} T{}^{\mu\nu}_s d\sigma{}_\nu = \mu_s\left(\delta{}^\mu_r - \frac{a}{r^2}\delta{}^\mu_\phi \right)
\ee
This results in a change of energy $E$ and angular momentum $J$ according to
\be
\dot{E} = j{}^\mu  \xi{}_\mu = 0 , \quad
\dot{J} = j{}^\mu \eta{}_\mu = -\mu_s a \sin^2\theta_0 .
\ee }

\begin{figure}
  \centering
  \includegraphics[width=0.3\textwidth]{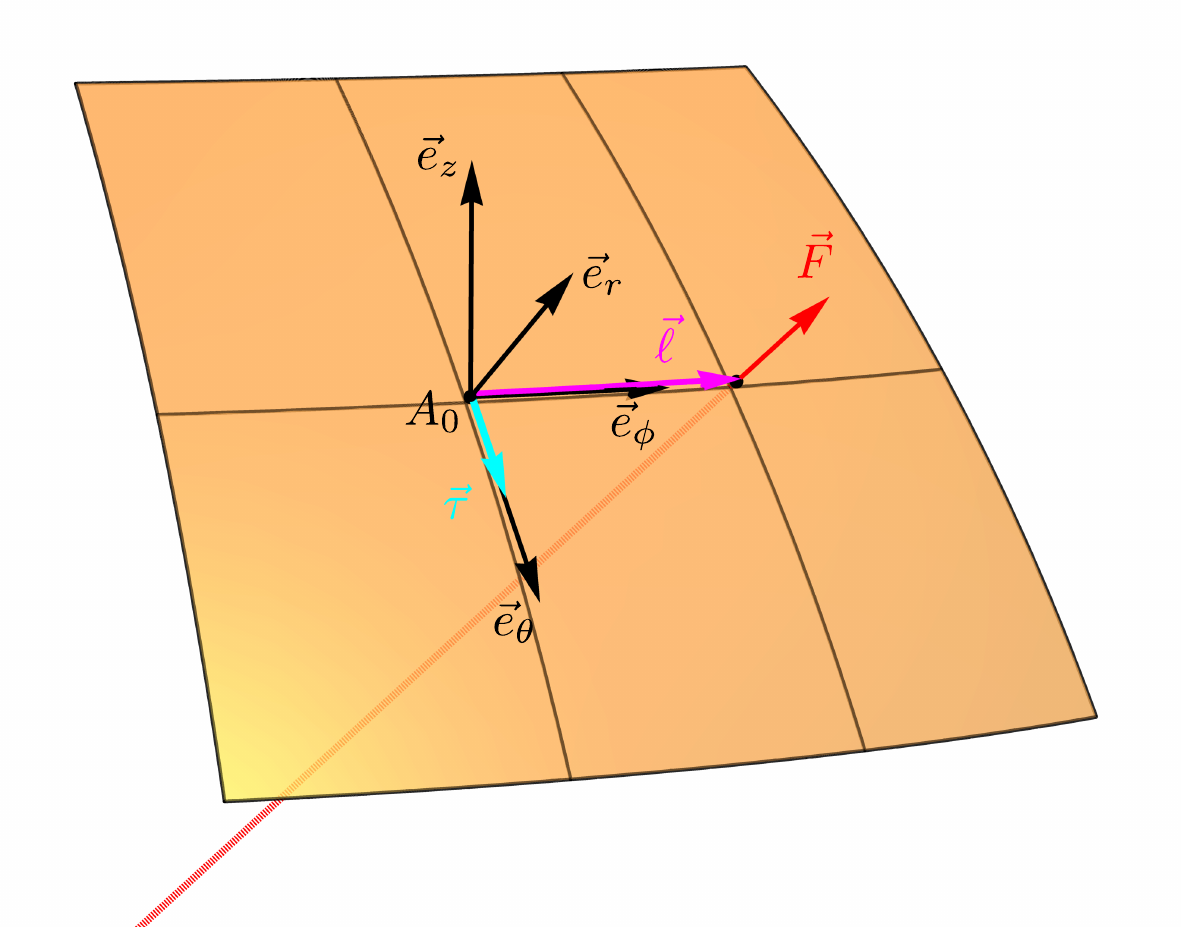}
  \caption{Illustration of the calculation of the torque. The picture shows a small piece of the sphere of large radius $r$. At a point $A_0$ with spherical coordinates $(\theta_0,\phi_0)$ one introduces a triad of unit orthonormal vectors $\{ \vec{e}_r,\vec{e}_{\theta},\vec{e}_{\phi}\}$. Two of these vectors, in the $\theta$ and $\phi$ directions, span a two-dimensional plane tangent to the sphere. Since the sphere's curvature is small this plane practically coincides with the sphere at $A_0$. The string enters the tangent plane orthogonally. Its direction coincides with $\vec{e}_r$ and its position is shifted from $A_0$ by the distance $\vec{\ell}$ in the direction of $\vec{e}_{\phi}$. The unit vector $\vec{e}_z$ is in the direction of the angular momentum of the black hole.
  }
  \label{fig:visualization-torque}
\end{figure}

\jen{Let us now interpret these results.} We introduce three mutually orthonormal vectors $\{ \vec{e}_r,\vec{e}_{\theta}, \vec{e}_{\phi}\}$ directed along $r$, $\theta$ and $\phi$ coordinate lines, respectively. For this choice the triad of the vectors has the right-hand orientation. The displacement \jen{$\vec{\ell}$} of the string position from the origin of the frame is $a\sin\theta_0$ in the \val{positive $\phi$}  direction,
\be
\vec{\ell}=\val{+}a\sin\theta_0 \vec{e}_{\phi}\, .
\ee
\jen{In order to keep the string in equilibrium one needs to apply a force along the string's spatial tangent vector, which asymptotically takes the form}
\be
\vec{F}=\mu_s \vec{e}_r\, .
\ee
\jen{In the frame at $A_0$} this force would provide the torque
\be
\vec{\tau}=\vec{\ell}\times \vec{F}=\mu_s a \sin\theta_0 \vec{e}_{\theta}\, .
\ee
A unit vector $\vec{e}_z$, which is parallel to the direction of the angular momentum of the black hole, is
\be
\vec{e}_z=\cos\theta_0 \vec{e}_r-\sin\theta_0 \vec{e}_{\theta}\, .
\ee
Thus, the projection of the torque on the direction of the angular momentum of the black hole is
\be
\tau_z=\val{-}\mu_s a \sin^2\theta_0 .
\ee
This torque decreases the angular momentum of the black hole, and the rate of this process is in agreement with the result (\ref{EJ})\footnote{\val{The force of the string on the black hole would also result in the motion of the black hole as a whole in the external space. In order to compensate this effect it is sufficient to attach an additional (``dual'') string with parameters $\theta_0'=\pi-\theta_0$ and $\phi_0'=\pi+\phi_0$. The action of such a dual string compensates the  force of the original one, while the loss of angular momentum would be doubled.}}.

\subsection{Rotating black holes with stringy hair}

It is rather straightforward to smear the string and consider a continuous distribution of the stringy matter around a rotating black hole. It is sufficient to use the approach developed in \cite{Ivanov:2001wb,Ivanov:2003sq}. In our case the description of a continuous ensemble of strings is greatly simplified because in the incoming coordinates $(v,r,\theta,\hat{\phi})$ the string looks like a straight object with $\theta$ and $\hat{\phi}$ constant on its world-sheet. Using this property we define the stress-energy tensor of the string distribution as follows
\be\n{aver}
T^{\mu\nu}=\int d\theta_0 d\hat{\phi}_0 \sin\theta_0 \mu_s^{-1}\mu(\theta_0,\hat{\phi}_0) T_s^{\mu\nu}\, ,
\ee
where the function $\mu(\theta,\hat{\phi})$ is the density distribution of the stringy matter. The result of the averaging (\ref{aver}) is
\be
T^{\mu\nu}=\mu \tau^{\mu\nu}\, .
\ee
The flux of the energy through the horizon vanishes, while the flux of angular momentum per unit time is
\be
\dot{J}=-a\int d\hat{\phi} d\theta \sin^2\theta \mu(\theta,\hat{\phi})\, .
\ee

\section{Principal Killing strings in Kerr--NUT--(A)dS spacetime}

\subsection{Principal Killing strings}

\val{We consider now} a generalization of a stationary single string solution which we discussed in the previous section. Namely, instead of the Kerr metric we consider its generalization. For this purpose we first rewrite the Kerr metric (\ref{Kerr}) in new coordinates
\be
\tau=t-a\phi\hhh y=a\cos\theta\hhh \psi=\phi/a\, .
\ee
Its form is
\ba \n{CAN}
&&ds^2=-{\Delta_r\over \Sigma} (d\tau+y^2 d\psi)^2+{\Delta_y\over \Sigma} (d\tau-r^2 d\psi)^2\nonumber\\
&&\quad\quad +{\Sigma\over \Delta_r} dr^2+{\Sigma\over \Delta_y} dy^2\, .
\ea
In the above, $\Sigma=r^2+y^2$. For the Kerr metric one has $\Delta_r=r^2-2Mr+a^2$ and $\Delta_y=a^2-y^2$.

In what follows we consider what is  called off-shell version of the canonical metric (\ref{CAN}). Namely, we assume that $\Delta_r$ and $\Delta_y$ are arbitrary functions of their arguments, $r$ and $y$, respectively. In particular, this means that in the general case the metric does not obey the vacuum Einstein equations. The properties of such metrics are discussed in \cite{frolov2011introduction,Frolov:2017kze}. For the special choice
\ba
&& \Delta_r=(r^2+a^2)(1-\Lambda r^2/3)-2Mr\, ,\\
&& \Delta_y=(a^2-y^2)(1+\Lambda y^2/3)+2Ny\, ,
\ea
the metric (\ref{CAN}) describes the Kerr--NUT--(A)dS black hole with $N$ being the NUT parameter.

In what follows we shall use the following results:
\begin{itemize}
\item The metric (\ref{CAN}) possesses the principal tensor $\ts{h}$, which is generated by the potential $\ts{b}$:
    \be
    \ts{h}=d\ts{b}\hh b=-{1\over 2}[(r^2-y^2) d\tau+r^2 y^2 d\psi]\, ;
    \ee
\item The Killing vector $\ts{\xi}=\partial_{\tau}$ is related to $\ts{h}$ as follows
\be
\xi^{\mu}={1\over 3} \nabla_{\nu} h^{\nu\mu}\, ;
\ee
\item The principal tensor $\ts{h}$ has four eigenvectors:
\ba
&&h^{\mu}_{\ \ \nu} l_{\pm}^{\nu}=\mp r l_{\pm}^{\mu}\, ,\\
&&h^{\mu}_{\ \  \nu} m_{\pm}^{\nu}=\pm iy m_{\pm}^{\mu}\, .
\ea
These eigenvectors can be written as follows:
\ba
&& l_{\pm}^{\mu}=\left( {r^2\over \Delta_r},\pm 1,0,{1\over \Delta_r}\right)\, ,\n{lll}\\
&& m_{\pm}^{\mu}=\left( {i y^2\over \Delta_y},0,\pm 1,-{i\over \Delta_y}\right)\, .
\ea
\item The null vectors $\ts{l}_{\pm}$ are generators of principal null geodesics in the affine parametrization
    \be
    l_{\pm}^{\nu} l_{\pm \ ;\nu}^{\mu}=0\, .
    \ee
\item The principal null vectors $\ts{l}_{\pm}$ are also eigenvectors of the 2-form $F_{\mu\nu}=\xi_{\mu ;\nu}$ constructed from the primary Killing vector $\ts{\xi}$:
    \be
    F^{\mu}_{\ \ \nu}  l_{\pm}^{\nu} =\pm \kappa l_{\pm}^{\mu}\hhh
    \kappa=\partial_r\left( {\Delta_r-\Delta_y\over 2\Sigma}\right)\, .
    \ee
\item Let us denote the Lie derivative along $\ts{\xi}$ by ${\cal L}_{\ts{\xi}}$. Then one has
\be
{\cal L}_{\ts{\xi}}\ts{l}_{\pm}= [\ts{\xi},\ts{l}_{\pm}]=0\hhh
{\cal L}_{\ts{\xi}}\ts{h}=0\, .
\ee
\end{itemize}
These relations can be  easily checked by using computer programs, e.g.\ GRTensor.

The stationary string equations in four and higher dimensional Kerr--NUT--(A)dS spacetime allow a complete separation of variables \cite{Carter:1989bs,Kubiznak:2007ca}. Here we consider a special case of a stationary string which regularly crosses the event horizon. We call this solution of the string equation a \emph{principal Killing string}.

To construct this solution we choose one of two null principal vector fields and denote it by $\ts{l}$ (without a subscript $\pm$)\footnote{In what follows, in our construction of the stationary string solution we choose the vector $\ts{l}=\ts{l}_-$, which is regular at the future event horizon.}.
Since two vectors $\ts{\xi}$ and $\ts{l}$ commute, according to the Frobenius theorem the spacetime is foliated by two-dimensional surfaces $\Sigma$, such that both of these vectors are tangent to it. There also  exist coordinates $x^{\mu}=(v,\lambda,y^i)$, $(i=2,3)$, such that for each $\Sigma$ \jen{one has} $y^i=\text{const}$ and $z^a=(v,\lambda)$, $(a=0,1)$, are coordinates on $\Sigma$ such that
\be
\ts{\xi}=\partial_v\hh \ts{l}=\partial_{\lambda}\, .
\ee

In the coordinates $x^{\mu}=(z^a,y^i)$ one has
\be
ds^2=d\gamma^2+b_{ij}dy^i dy^j\, ,
\ee
where $d\gamma^2$ is the induced geometry on $\Sigma$,
\be
d\gamma^2=\gamma_{ab} dz^a dz^b=\ts{\xi}^2 dv^2 +2 (\ts{\xi},\ts{l}) dv d\lambda\, .
\ee
Denote by $n_{(i)}^{\mu}$ two mutually orthogonal unit normal vectors to $\Sigma$. An extrinsic curvature of $\Sigma$ is
\be
\Omega_{(i) ab}=g_{\mu\nu} n^{\mu}_{(i)} x^{\rho}_{,a}\nabla_{\rho} x^{\nu}_{,b}\, .
\ee
The surface $\Sigma$ is minimal if the following conditions are valid
\ba
&&\Omega_{(i)}\equiv \gamma^{ab}\Omega_{(i) ab}=g_{\mu\nu} n_{(i)}^{\mu} Z^{\nu}=0\, ,\\
&& Z^{\nu}=\gamma^{ab}x^{\rho}_{,a}\nabla_{\rho} x^{\nu}_{,b}\, .
\ea
Using the relations
\be
\partial^2_{\gamma}={2\over (\ts{\xi},\ts{l})}\partial_v \partial_{\lambda}-{\ts{\xi}^2 \over (\ts{\xi},\ts{l})^2}\partial_{\lambda}^2\hhh x^{\mu}_{,v}=\xi^{\mu}\hhh x^{\mu}_{,\lambda}=l^{\mu}\, ,
\ee
one gets
\be
Z^{\nu}={1\over (\ts{\xi},\ts{l})}\left( \xi^{\rho}\nabla_{\rho} l^{\nu}
+ l^{\rho}\nabla_{\rho} \xi^{\nu}\right)-{\ts{\xi}^2\over (\ts{\xi},\ts{l})^2}l^{\rho}\nabla_{\rho}l^{\nu}\, .
\ee
The properties of $\ts{l}$ and $\ts{\xi}$, mentioned above, \jen{imply that $Z^{\nu}\sim l{}^\nu$}. Hence the two-surface generated by these vectors is minimal. We call it \emph{principal Killing surface}. Since one of the tangent vectors, $\ts{l}$, is null, the minimal surface $\Sigma$ is timelike and represents a special time independent solution of the Nambu--Goto equations. We call such strings \emph{principal Killing strings}.

\subsection{Principal Killing strings in the incoming null coordinates}

Let us change coordinates \jen{in \eqref{CAN} to}
\be
d\tau =dv -{r^2\over \Delta_r} dr -a d\hat{\phi}\hhh
d\psi=a^{-1} d\hat{\phi} -{dr\over \Delta_r}\, .
\ee
The off-shell canonical metric in these new coordinates $(v,r,y,\hat{\phi})$ is then given by
\ba\n{off}
&&ds^2=-{\Delta_r\over \Sigma} \left(dv-{a^2-y^2\over a} d\hat{\phi}\right)^2
+{\Delta_y\over \Sigma} \left(dv-{r^2+a^2\over a} d\hat{\phi}\right)^2\nonumber\\
&&\quad\quad +2 \left(dv-{a^2-y^2\over a} d\hat{\phi}\right) dr +{\Sigma\over \Delta_y} dy^2\, ,
\ea
\jen{such that $\sqrt{-g} = \Sigma/a$}. The incoming principal null vector $\ts{l}$, see Eq.~(\ref{lll}), takes the form \jen{$\ts{l}=-\partial_r$}. Hence one can identify the affine parameter $\lambda$ with the coordinate $r$. One also has
\ba
&&\ts{\xi}^2={\Xi\over \Sigma}\hh (\ts{\xi},\ts{l})=\jen{-1}\, ,\\
&&\Xi=\Delta_y-\Delta_r\, ,\n{XXX}
\ea
so that the induced metric on the surface of the principal Killing string in the metric (\ref{off}) is
\be
d\gamma^2={\Xi\over \Sigma} dv^2 \jen{+2dv dr} \, .
\ee

The string  in the incoming coordinates is ``straightened'', so that $\theta$ and $\hat{\phi}$ are constant on its surface. For this reason the calculation of the stress-energy tensor for such a string is straightforward and \jen{can be simply obtained by repeating the calculations for the Kerr metric:
\ba
&&T_s^{\mu\nu}=q \tau^{\mu\nu}\, ,\  \tau^{\mu\nu}=\jen{ {1 \over \Sigma} \left( \jen{2\delta_v^{(\mu}\delta_r^{\nu)}}-{\Xi\over \Sigma}\delta_r^{\mu}\delta_r^{\nu}\right)\, ,}\\
&&q=q(y,\hat{\phi}|y_0,\hat{\phi}_0)=\mu_s a \delta(y-y_0)\delta(\hat{\phi}-\hat{\phi}_0)\, .
\ea
Due to $d y = -\sin\theta \, d\theta$ the horizon surface element is
\ba
&&d\sigma_{\mu}\hor + \frac{\Sigma_+}{a} \delta_{\mu}^r dv dy d\hat{\phi} \, .
\ea
The flux vector is hence given by
\be
j^{\mu}\hor \mu_s \left(- \delta_v^{\mu}+{\Delta_y \over \Sigma_H} \delta_r^{\mu}\right)\, .
\ee }
The fluxes of the energy and angular momentum through the horizon are
\be
\dot{E}\hor  \xi_{\mu} j^{\mu} = 0\hhh
\dot{J}\hor \eta_{\mu} j^{\mu} = -\frac{\mu_s}{a} \Delta_{y_0} \, .
\ee

\jen{In general, the off-shell metric \eqref{off} is not asymptotically flat, which is why a discussion of asymptotic properties is not well-defined. We omit this here.}

\jen{However, by repeating the arguments presented above, we can define smeared principal Killing string matter:
\be\n{smeared-killing}
T^{\mu\nu}= \int dy_0 d\hat{\phi}_0 \mu_s^{-1}\mu(y_0,\hat{\phi}_0) T_s^{\mu\nu}\, ,
\ee
such that the flux of angular momentum turns out to be
\be
\dot{J} = - \frac{1}{a} \int dy \, d\hat{\phi} \, \Delta_y \, \mu(y,\psi)\, .
\ee
 }

\section{Discussion}

Strings that pierce a black hole can be used for effective quantum energy mining from them. When the number of strings becomes large, one can approximate their distribution by stringy matter. One might say that such black holes have stringy hair. For static black holes the stringy matter deforms the black hole geometry, described by a warped metric. This metric is a direct sum of the $(t,r)$ sector of the original solution of the corresponding Einstein equations, while the warped part is a deformed unit 2-sphere. The Gaussian curvature $K=1+\Phi$ differs from the Gaussian curvature of a round unit sphere by a term $\Phi$ proportional to the energy density of the stringy matter.

The case when the black hole is rotating is much more complicated. The reason is that if a single string is attached to the black hole, it produces a torque which permanently decreases the angular momentum of the black hole. Correspondingly, the stringy matter has a similar effect, and we calculated the rate of the loss of the angular momentum of the black hole for both cases. Moreover, we obtained a new solution for a stationary string in the spacetime of the Kerr--NUT--(A)dS black hole which we call a \emph{principal Killing string}. It has the property that both the primary Killing vector and the tangent vector to the principal null rays are tangent to it. Also, this string configuration is regular in the vicinity of the event horizon. We calculated the flux of the angular momentum from such a black hole in the cases of a single string and stringy matter. Unfortunately, the backreaction problem is rather difficult in this case{\jen{: this is caused by the time-dependency of the the geometry including the backreaction effects.} The study of the evolution of rotating black holes interacting with cosmic strings is a quite interesting problem.

\section*{Acknowledgments}
\jen{We thank our anonymous referee for pointing out the relation of Eq.~\eqref{vvv} to the class of Robinson--Trautman spacetimes.} \jen{J.B.\ is grateful for a Vanier Canada Graduate Scholarship administered by the Natural Sciences and Engineering Research Council of Canada as well as for the Golden Bell Jar Graduate Scholarship in Physics by the University of Alberta.}
V.F.\ thanks the Natural Sciences and Engineering Research Council of Canada and the Killam Trust for financial support.

\bibliography{STRINGY_BHS_3}{}

\end{document}